\documentclass[12pt]{article}
\makeatletter
\newcommand{\singlespacing}{\let\CS=\@currsize\renewcommand{\baselinestretch}{1}\tiny\CS}
\usepackage{epsfig}

\usepackage{graphicx}
\usepackage{verbatim}   
\usepackage{color}      
\usepackage{subfigure}  
\usepackage{amsfonts}
\usepackage{array,multirow}

\begin{document}
\baselineskip=24pt
\parskip = 10pt
\def \qed {\hfill \vrule height7pt width 5pt depth 0pt}
\newcommand{\ve}[1]{\mbox{\boldmath$#1$}}
\newcommand{\IR}{\mbox{$I\!\!R$}}
\newcommand{\1}{\Rightarrow}
\newcommand{\bs}{\baselineskip}
\newcommand{\esp}{\end{sloppypar}}
\newcommand{\be}{\begin{equation}}
\newcommand{\ee}{\end{equation}}
\newcommand{\beanno}{\begin{eqnarray*}}
\newcommand{\inp}[2]{\left( {#1} ,\,{#2} \right)}
\newcommand{\eeanno}{\end{eqnarray*}}
\newcommand{\bea}{\begin{eqnarray}}
\newcommand{\eea}{\end{eqnarray}}
\newcommand{\ba}{\begin{array}}
\newcommand{\ea}{\end{array}}
\newcommand{\nno}{\nonumber}
\newcommand{\dou}{\partial}
\newcommand{\bc}{\begin{center}}
\newcommand{\ec}{\end{center}}
\newcommand{\2}{\subseteq}
\newcommand{\cl}{\centerline}
\newcommand{\ds}{\displaystyle}
\newcommand{\mr}{\mathbb{R}}
\newcommand{\mn}{\mathbb{N}}

\def\refhg{\hangindent=20pt\hangafter=1}
\def\refmark{\par\vskip 2.50mm\noindent\refhg}

\title{\sc Multivariate Geometric Skew-Normal Distribution}

\author{Debasis Kundu$^{1}$}

\date{}
\maketitle

\begin{abstract}
Azzalini \cite{Azzalini:1985} introduced a skew-normal distribution of which normal distribution is a special case.  Recently Kundu 
\cite{Kundu:2014} introduced a geometric skew-normal distribution and showed that it has certain advantages over Azzalini's 
skew-normal distribution.
In this paper we discuss about the multivariate geometric skew-normal distribution.  It can be used as an alternative to Azzalini's 
skew normal distribution.  We discuss different properties of the proposed distribution.  It is observed that the joint probability
density function of the multivariate geometric skew normal distribution can take variety of shapes. Several characterization 
results have been established.  Generation from a multivariate geometric skew normal distribution is quite simple, hence the 
simulation experiments can be performed quite easily.  The maximum likelihood estimators of the
unknown parameters can be obtained quite conveniently using expectation maximization (EM) algorithm.  We perform some simulation 
experiments and it is observed that the performances of the proposed EM algorithm are quite satisfactory.
Further, the analyses of two data sets have been performed, and it is 
observed that the proposed methods and the model work very well.  

\end{abstract}



\noindent$^1$Department of Mathematics and Statistics, Indian Institute of Technology Kanpur, Kanpur, 
Pin 208016, India.  \ \ e-mail: kundu@iitk.ac.in.

\newpage

\section{\sc Introduction}

Azzalini \cite{Azzalini:1985} proposed a class of three-parameter skew-normal distributions which includes the normal one.  
Azzalini's skew normal 
(ASN) distribution has received a considerable attention in the last two decades due to its flexibility and its applications in different 
fields.  The probability density function (PDF) of ASN takes the following form:
$$
f(x; \mu,\sigma,\lambda) = \frac{2}{\sigma} \phi \left ( \frac{x-\mu}{\sigma} \right ) \Phi \left ( \frac{\lambda(x-\mu)}{\sigma} 
\right ), \ \ \ \ -\infty < x, \mu, \lambda < \infty, \ \  \sigma > 0,   
$$
where $\phi(x)$ and $\Phi(x)$ denote the standard normal PDF and standard normal cumulative distribution function (CDF), respectively, 
at the point $x$.  Here $\mu$, $\sigma$ and $\lambda$ are known as the location, scale and skewness or tilt parameters, respectively.  ASN 
distribution has an unimodal PDF, and it can be both positively or negatively skewed depending on the skewness parameter.  Arnold and
Beaver \cite{AB:2000}  provided an interesting interpretation of this model in terms of hidden truncation.  This model has been used quite 
effectively to analyze skewed data in different fields due to its flexibility.

Later Azzalini and Dalla Valle \cite{AD:1996} constructed a multivariate distribution
with skew normal marginals.   From now on we call it as Azzalini's multivariate skew-normal (AMSN) distribution, and it can be defined
as follows.  A random vector ${\ve Z} = (Z_1, \ldots, Z_d)^T$ is a $d$-dimensional AMSN distribution, if it has the following PDF
$$
g({\ve z}) = 2 \phi_d({\ve z}; {\ve \Omega}) \Phi({\ve \alpha}^T {\ve z}), \ \ \  {\ve z} \in \mr^d,  
$$
where $\phi_d({\ve z}, {\ve \Omega})$ denotes the PDF of the $d$-dimensional multivariate normal distribution with standardized 
marginals, and correlation matrix ${\ve \Omega}$.  We denote such a random vector as ${\ve Z} \sim$ 
SN$_d({\ve \Omega}, {\ve \alpha})$.  Here the vector ${\ve \alpha}$ is known as the shape vector, and it can be easily seen
that the PDF of AMSN distribution is unimodal and can take different shapes.  It has several interesting properties, and it has
been used quite successfully to analyze several multivariate data sets in different areas because of its flexibility.

Although ASN distribution is a very flexible distribution, it cannot be used to model moderate or heavy tail data; see for example 
Azzalini and Capitanio \cite{AC:2014}.  It is well known to be a 
thin tail distribution.  Since the marginals of 
AMSN are ASN, multivariate heavy tail data cannot be modeled by using AMSN.  Due to this reason several other skewed distributions, 
often called skew-symmetric distributions, have been suggested in the literature using different kernel functions other than the 
normal kernel function and using the same technique as Azzalini \cite{Azzalini:1985}.  Depending on the kernel function the resulting distribution
can have moderate or heavy tail behavior.  Among different such distributions, skew-t distribution is quite commonly used in practice, 
which can produce heavy tail distribution depending on the degrees of freedom of the associated $t$-distribution.  It has a multivariate extension also.  For a detailed discussions on different skew-symmetric distribution, the readers are referred to the excellent monograph
by Azzalini and Capitanio \cite{AC:2014}.

Although ASN model is a very flexible one dimensional model, and it has several interesting properties, it is well known that 
computing the maximum likelihood estimators (MLEs) of the unknown parameters of an ASN model is a challenging issue. Azzalini 
\cite{Azzalini:1985} has shown that
there is a positive probability that the MLEs of the unknown parameters of a ASN model do not exist.  If all the data points 
have same sign, then the MLEs of unknown parameters of the ASN model may not exist.  The problem becomes more 
severe for AMSN model, and the problem exists for other kernels also.

Recently, the author \cite{Kundu:2014} proposed a new three-parameter skewed distribution, of which normal distribution is a 
special case.  The proposed distribution can be obtained as a geometric sum of independent identically distributed (i.i.d.) normal
random variables, and it is called as the geometric skew normal (GSN) distribution.  It can be used quite effectively as an 
alternative to an ASN distribution. It is observed that the GSN distribution is a very flexible distribution, as its PDF can take different
shapes depending on the parameter values.  Moreover, the MLEs of the unknown parameters can be computed quite conveniently using
the EM algorithm.  It can be easily shown that the `pseudo-log-likelihood' function has a unique maximum, and it can be obtained in
explicit forms.  Several interesting properties of the GSN distribution have also been developed by Kundu \cite{Kundu:2014}.

The main aim of this paper is to consider the multivariate geometric skew-normal (MGSN) distribution, develop its  various properties  
and discuss different inferential issues.  Several 
characterization results and dependence properties have also been established.  It is observed that the generation from a MGSN 
distribution is quite simple, hence simulation experiments can be performed quite conveniently.  Note that the
$d$-dimensional MGSN model has $d+1 + d(d+1)/2$ unknown parameters.  The MLEs of the unknown parameters can be obtained by solving 
$d+1 + d(d+1)/2$ non-linear equations.   We propose to use EM algorithm, and it is observed that the 'pseudo-log-likelihood' 
function has a unique maximum, and it can be obtained in explicit forms.  Hence, the implementation of the EM algorithm is quite  
simple, and the algorithm is very efficient.  We perform some simulation experiments to see the performances of the proposed EM 
algorithm and the performances are quite satisfactory.  We also perform the analyses of two data sets to illustrate how the 
proposed methods can be used in practice.  It is observed that the proposed methods and the model work quite satisfactorily. 

The main motivation to introduce the MGSN distribution can be stated as follows.  Although there are several skewed distributions 
available in one-dimension, the same is not true in $\mathbb{R}^d$.  The proposed MGSN distribution is a very flexible multivariate 
distribution which can 
produce variety of shapes.  The joint PDF can be unimodal or multimodal and the marginals can have heavy tails depending on the 
parameters.  It has
several interesting statistical properties.  Computation of the MLEs can be performed in a very simple manner even in high dimension.
Hence, if it is known that the data are obtained from a multivariate skewed distribution, the proposed model can be used for 
analysis purposes.  Generating random samples from a MGSN 
distribution is quite simple, hence any simulation experiment related to this distribution can be performed quite conveniently.  
Further, it is observed that in one of our data example the MLEs of AMSN do not exist, whereas the MLEs of MGSN distribution exist.
Hence, in certain cases the implementation of MGSN distribution becomes easier than the AMSN distribution. 
The 
proposed MGSN distribution provides a choice to a practitioner of a new multivariate skewed distribution to analyze multivariate 
skewed data.

Rest of the paper is organized as follows.  In Section 2, first we briefly describe the univariate GSN model, and discuss some of
its properties, and then we describe MGSN model.  Different properties are discussed in Section 3.  In Section 4, we discuss the 
implementation of the EM algorithm, and some testing of hypotheses problems.  Simulation results are presented in Section 5.  
The analysis of two data sets are presented in Section 6, and finally we conclude the paper in Section 7.

\section{\sc GSN and MGSN Distributions}

We use the following notations in this paper.  A normal random variable with mean $\mu$ and variance $\sigma^2$ will be denoted
by N$(\mu, \sigma^2)$.  A $d$-variate normal random variable with mean vector ${\ve \mu}$ and dispersion matrix ${\ve \Sigma}$
will be denoted by N$_d({\ve \mu}, {\ve \Sigma})$.  The corresponding PDF and CDF at the point ${\ve x}$ will be denoted by 
$\phi_d({\ve x}; {\ve \mu}, {\ve \Sigma})$ and $\Phi_d({\ve x}; {\ve \mu}, {\ve \Sigma})$, respectively.  A geometric random variable
with parameter $p$ will be denoted by GE$(p)$, and it has the probability mass function (PMF): $p(1-p)^{n-1}$ for 
$n = 1, 2, \ldots.$

\subsection{\sc GSN Distribution} 

Suppose $N \sim$ GE$(p)$ and $\{X_i; i = 1, 2, \ldots, \}$ are i.i.d. Gaussian random variables.  It is assumed that $N$ and $X_i$'s are 
independently distributed.  Then the random variable 
$$
X \stackrel{dist}{=} \sum_{i=1}^N X_i
$$
is known as GSN random variable and its distribution will be denoted by GSN$(\mu,\sigma,p)$.  Here, `$\ds  \stackrel{dist}{=}$' means equal in distribution.   The GSN 
distribution can be seen as one of the compound geometric distributions.
The PDF of $X$ takes the following form:
$$
f_X(x; \mu,\sigma,p) = \sum_{k=1}^{\infty} \frac{p}{\sigma \sqrt{k}} \phi \left ( \frac{x-k \mu}{\sigma \sqrt{k}} \right ) (1-p)^{k-1}.
$$
When $\mu$ = 0 and $\sigma$ = 1, we say that $X$ has a standard GSN distribution, and it will be denoted by GSN$(p)$.

The standard GSN is symmetric about 0, and unimodal, but the PDF of GSN$(\mu,\sigma,p)$ can take different shapes.  It can be 
unimodal or multimodal depending on $\mu$, $\sigma$ and $p$ values.  The hazard function is always an increasing function.  If $X \sim$ 
GSN$(\mu,\sigma,p)$, then the moment generating function (MGF) of $X$ becomes
\be
M_X(t) = \frac{p e^{\mu t + \frac{\sigma^2 t^2}{2}}}{1 - (1-p) e^{\mu t + \frac{\sigma^2 t^2}{2}}}, \ \ \  t \in A_1(\mu,\sigma,p),    \label{mgf-gsn}
\ee
where 
\beanno
A_1(\mu, \sigma, p) & = &\left \{t; t\in \mr, (1-p) e^{\mu t + \frac{\sigma^2 t^2}{2}} < 1 \right \}  \\
& =  &\left \{t; t\in \mr, 2 \mu t + \sigma^2 t^2 + 2 \ln (1-p) < 0 \right \}.
\eeanno
The corresponding cumulant generating (CGF) function of $X$ is
\be
K_X(t) = \ln M_X(t) = \ln p + \mu t + \frac{\sigma^2 t^2}{2} - \ln \left (1 - (1-p) e^{\mu t + \frac{\sigma^2 t^2}{2}} \right ).
\label{cgf-gsn} 
\ee
From (\ref{cgf-gsn}), the mean, variance, skewness and kurtosis  can be easily obtained as
\bea
E(X) & = & \frac{\mu}{p},  \label{mean-gsn}    \\
V(X) & = & \frac{\sigma^2 p + \mu^2(1-p)}{p^2},   \label{var-gsn} \\      
\gamma_1 & = & \frac{(1-p) \left (\mu^3(2-p) + 3 \mu \sigma^2 p \right )}{(p \sigma^2 + \mu^2(1-p))^{3/2}},  \nonumber  \\
\gamma_2 & = & \frac{\mu^4(1-p)(p^2 - 6p + 6) - 2\mu^2 \sigma^2p(1-p)(p^2+3p-6) + 3 \sigma^4 p^2}{(p \sigma^2 + \mu^2(1-p))^2},
\nonumber
\eea
respectively.  It is clear from the expressions of (\ref{mean-gsn}) and (\ref{var-gsn}) that as $p \rightarrow 0$, $|E(X)|$ and $V(X)$ diverge to $\infty$.  It 
indicates that GSN model can be used to model heavy tail data.  It has been shown that the GSN law is infinitely divisible, and 
an efficient EM algorithm has been suggested to compute the MLEs of the unknown parameters.  

\subsection{\sc MGSN Distribution}
 A $d$-variate MGSN distribution can be defined as follows.  Suppose $N \sim$ GE$(p)$, $\{{\ve X}_i; i = 1,2 \ldots\}$ are i.i.d.
N$_d({\ve \mu}, {\ve \Sigma})$ random vectors and all the random variables are independently distributed.  Define
\be
{\ve X} \stackrel{dist}{=} \sum_{i = 1}^N {\ve X}_i,   \label{mgsn-def}
\ee
then ${\ve X}$ is said to have a $d$-variate geometric skew-normal distribution with parameters $p$, ${\ve \mu}$ and ${\ve \Sigma}$,
and its distribution  will be denoted by MGSN$_d(p, {\ve \mu}, {\ve \Sigma})$.  If ${\ve X} \sim$ MGSN$_d(p,{\ve \mu}, {\ve \Sigma})$, then 
the CDF and PDF of ${\ve X}$ become
$$
F_{\ve X}({\ve x}; {\ve \mu}, {\ve \Sigma}, p) = \sum_{k=1}^{\infty} p(1-p)^{k-1} \Phi_d({\ve x}; k {\ve \mu}, k {\ve \Sigma}) 
$$
and
\beanno
f_{\ve X}({\ve x}; {\ve \mu}, {\ve \Sigma}, p) & = & \sum_{k=1}^{\infty} p(1-p)^{k-1} \phi_d({\ve x}; k {\ve \mu}, k {\ve \Sigma})   
\nonumber \\
& = & \sum_{k=1}^{\infty} \frac{p(1-p)^{k-1}}{(2 \pi)^{d/2} |{\ve \Sigma}|^{1/2} k^{d/2}}
e^{-\frac{1}{2k} ({\ve x} - k {\ve \mu})^T {\ve \Sigma}^{-1}({\ve x} - k {\ve \mu})},
\eeanno
respectively.  Here $\ds \Phi_d({\ve x}; k {\ve \mu}, k {\ve \Sigma})$ and $\ds \phi_d({\ve x}; k {\ve \mu}, k {\ve \Sigma})$ 
denote the CDF and PDF of a $d$-variate normal distribution, respectively,  with the mean vector $\ds k {\ve \mu}$ and 
dispersion matrix $\ds k {\ve \Sigma}$.

If ${\ve \mu} = {\ve 0}$ and ${\ve \Sigma} = {\ve I}$, we say that ${\ve X}$ is a standard $d$-variate MGSN random variable, 
and its distribution  will be denoted by MGSN$_d(p)$.  The PDF of MGSN$_d(p)$ is symmetric and unimodal, for all values of $d$ and $p$,
whereas the PDF of MGSN$_d(p,{\ve \mu}, {\ve \Sigma})$ may not be symmetric, and it can be unimodal or multimodal depending
on parameter values.  The MGF of MGSN can be obtained in explicit form.  If ${\ve X} \sim$ MGSN$_d(p, {\ve \mu}, {\ve \Sigma})$, then the MGF of ${\ve X}$ is
\be
M_{\ve X}({\ve t}) = 
 \frac{p e^{{\ve \mu}^T {\ve t} + \frac{1}{2} {\ve t}^T {\ve \Sigma} {\ve t}}}{1 - (1-p)  e^{{\ve \mu}^T {\ve t} + \frac{1}{2} {\ve t}^T {\ve \Sigma} {\ve t}}}, \ \ \ \
 {\ve t} \in A_d({\ve \mu}, {\ve \Sigma}, p),    \label{mgf-mgsn}
\ee
where
\beanno
 A_d({\ve \mu}, {\ve \Sigma}, p) & = & \left \{{\ve t}; {\ve t} \in  \mr^d, (1-p)  
e^{{\ve \mu}^T {\ve t} + \frac{1}{2} {\ve t}^T {\ve \Sigma} {\ve t}} < 1 \right \}  \\
& = & \left \{ {\ve t}; {\ve t} \in  \mr^d, {\ve \mu}^T {\ve t} + \frac{1}{2} {\ve t}^T {\ve \Sigma} {\ve t} + \ln (1-p) < 0 \right \}.
\eeanno
Further the generation of MGSN distribution is very simple.  The following algorithm can be used to generate samples from a 
MGSN random variable.


\noindent {\sc Algorithm 1:}
\begin{itemize}
\item Step 1: Generate $n$ from a GE$(p)$
\item Step 2: Generate ${\ve X} \sim$ N$_d(n {\ve \mu}, n {\ve \Sigma})$. 
\end{itemize}

\begin{figure}[h]
\bc
\epsfig{file=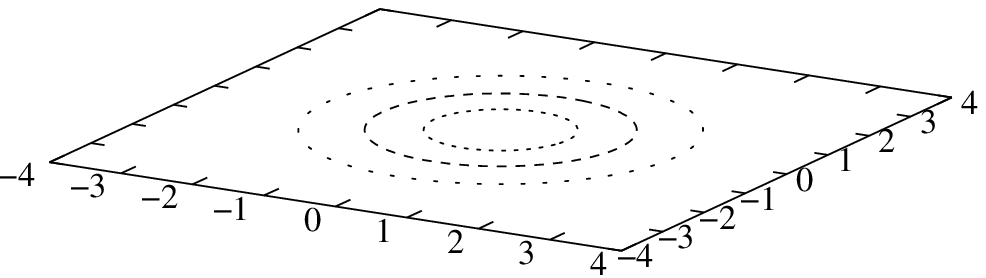, scale = 1.0}
\epsfig{file=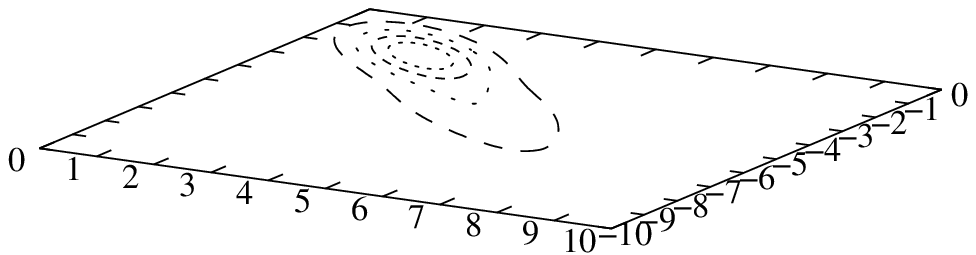, scale = 1.0}  \\
\epsfig{file=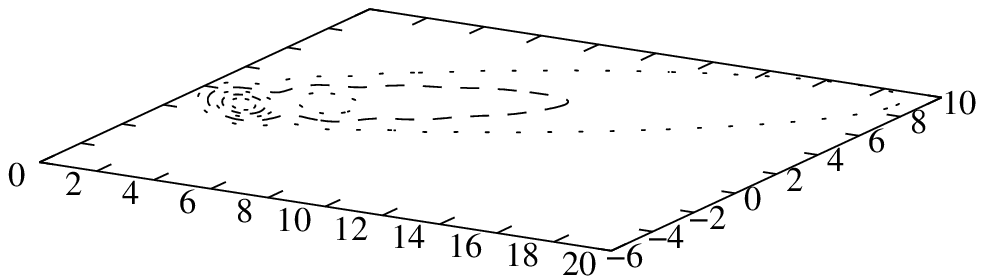, scale = 1.0}
\epsfig{file=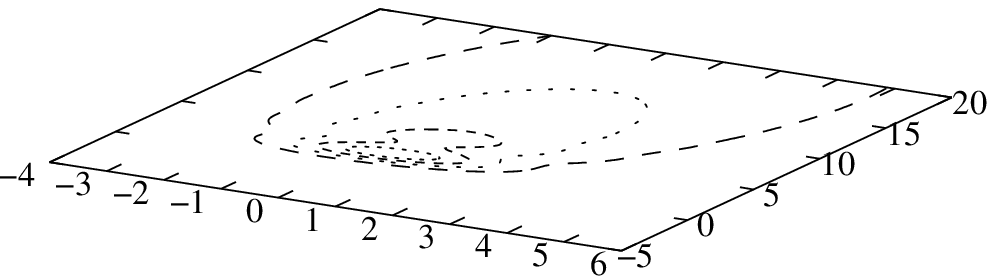, scale = 1.0}
\caption{The joint PDF of a bivariate geometric skew normal distribution for different parameter values.
    \label{jpdf}}
\ec
\end{figure}

In Figure \ref{jpdf} we provide the joint PDF of a bivariate geometric skew normal distribution for different parameter values:
(a) $p$ = 0.75, $\mu_1 = \mu_2$ = 0, $\sigma_1^2 = \sigma_2^2$ = 2, $\sigma_{12} = \sigma_{21} = 0$,
(b) $p$ = 0.5, $\mu_1 = \mu_2$ = 2.0, $\sigma_1^2 = \sigma_2^2$ = 1, $\sigma_{12} = \sigma_{21} = -0.5$,  
(c) $p$ = 0.15, $\mu_1$ = 2.0, $\mu_2$ = 1.0, $\sigma_1^2 = \sigma_2^2$ = 1, $\sigma_{12} = \sigma_{21} = -0.5$,  
(d) $p$ = 0.15, $\mu_1$ = 0.5, $\mu_2$ = -2.5, $\sigma_1^2 = \sigma_2^2$ = 1.0, $\sigma_{12} = \sigma_{21} = 0.5$.  

\section{\sc Properties}

In this section we discuss different properties of a MGSN distribution.  We use the following notations:
\be
{\ve X} = \left ( \matrix{{\ve X}_1 \cr {\ve X}_2 \cr} \right ),  \ \ {\ve \mu} = \left ( \matrix{{\ve \mu}_1 \cr {\ve \mu}_2 \cr} \right ), \ \ \ {\ve \Sigma} = \left ( \matrix{{\ve \Sigma}_{11} & {\ve \Sigma}_{12} \cr {\ve \Sigma}_{21} & 
{\ve \Sigma}_{22}  \cr} \right ).      \label{partition}
\ee
Here the vectors ${\ve X}$ and ${\ve \mu}$ are of the order $d$ each, and the matrix ${\ve \Sigma}_{11}$ is of the order 
$h \times h$.  Rest of the quantities are defined, so that they are compatible.  The following result provides the marginals
of a MGSN distribution.

\noindent {\sc Result 1:} If $\ds {\ve X} \sim \hbox{MGSN}_d(p, {\ve \mu}, {\ve \Sigma})$ and 
$\ds {\ve X}_1 \sim \hbox{MGSN}_h(p, {\ve \mu}_1, {\ve \Sigma}_{11})$ then
$$
{\ve X}_2 \sim MGSN_{d-h}(p, {\ve \mu}_2, {\ve \Sigma}_{22}). 
$$
\noindent {\sc Proof:} The result easily follows from the MGF of MGSN as provided in (\ref{mgf-mgsn}).   \qed

\noindent We further have the following results similar to the multivariate normal distribution.  The result may be used for 
testing simultaneously a set of linear hypothesis on the parameter vector ${\ve \mu}$ or it may have some independent interest 
also; see for example Rao \cite{Rao:1973}.

\noindent {\sc Theorem 1:} If ${\ve X} \sim$ MGSN$_d(p,{\ve \mu}, {\ve \Sigma})$, then ${\ve Z} = {\ve D} {\ve X} \sim$
MGSN$_s(p, {\ve D}{\ve \mu}, {\ve D} {\ve \Sigma} {\ve D}^T)$, where ${\ve D}$ is a $s \times d$ matrix of rank $s \le d$.

\noindent {\sc Proof:} The MGF of the random vector ${\ve Z}$ is
\beanno
M_{\ve Z}({\ve t}) & = & E \left ( e^{{\ve t}^T {\ve Z}} \right ) = E \left ( e^{{\ve t}^T {\ve D}{\ve X}} \right ) = 
E \left ( e^{ \left ({\ve D}^T {\ve t} \right )^T{\ve X}} \right )   \\
& = & \frac{p e^{\left ( {\ve D}{\ve \mu} \right )^T {\ve t} + \frac{1}{2} {\ve t}^T {\ve D}{\ve \Sigma} {\ve D}^T{\ve t}}}{1 - (1-p)  
e^{\left ( {\ve D}{\ve \mu} \right )^T {\ve t} + \frac{1}{2} {\ve t}^T {\ve D}{\ve \Sigma} {\ve D}^T{\ve t}}}, \ \ \ \ \hbox{for} \ \ \
 {\ve t} \in A_s^{D},
\eeanno
where
\beanno
A_s^D & = & \left \{{\ve t}; {\ve t} \in \mr^s, (1-p)  
e^{\left ( {\ve D}{\ve \mu} \right )^T {\ve t} + \frac{1}{2} {\ve t}^T {\ve D}{\ve \Sigma} {\ve D}^T{\ve t}} < 1 \right \}  \\
& = & \left \{{\ve t}; {\ve t} \in \mr^s, \ln (1-p) + 
\left ( {\ve D}{\ve \mu} \right )^T {\ve t} + \frac{1}{2} {\ve t}^T {\ve D}{\ve \Sigma} {\ve D}^T{\ve t} < 0 \right \}.
\eeanno
Hence the result follows.  \qed

If ${\ve X} = (X_1, \ldots, X_d)^T \sim$ MGSN$_d(p,{\ve \mu}, {\ve \Sigma})$, and if we denote ${\ve \mu}^T = 
(\mu_1, \ldots, \mu_d)$, ${\ve \Sigma} = ((\sigma_{ij}))$,  then the moments and cumulants 
of ${\ve X}$, for $i,j = 1, 2, \ldots, d$, can be obtained from  the MGF as follows:
\be
\left . E(X_i) =  \frac{\partial}{\partial t_i} M_{\ve X}({\ve t}) \right |_{{\ve t} = {\ve 0}} = \frac{\mu_i}{p}   \label{mgsn-mean}
\ee
$$
\left . E(X_i X_j)  =  \frac{\partial^2}{\partial t_i \partial t_j} M_{\ve X}({\ve t}) \right |_{{\ve t} = {\ve 0}} 
= \frac{p \sigma_{ij} + \mu_i \mu_j (2-p)}{p^2}.
$$
Hence, 
\be
Var(X_i) =  \frac{p \sigma_{ii} + \mu_i^2 (1-p)}{p^2},   \label{mgsn-var}
\ee
\be
Cov(X_i,X_j) = \frac{p \sigma_{ij} + \mu_i \mu_j (1-p)}{p^2},   \label{mgsn-cov}
\ee
and
\be
Corr(X_i, X_j) = \frac{p \sigma_{ij} + \mu_i \mu_j (1-p)}{\sqrt{p \sigma_{ii} + \mu_i^2(1-p)} \sqrt{p \sigma_{jj} + \mu_j^2(1-p)}}.
    \label{corr-mgsn}
\ee
It is clear from (\ref{corr-mgsn}) that the correlation between $X_i$ and $X_j$ for $i \ne j$, not only depends on $\sigma_{ij}$,
but it also depends on $\mu_i$ and $\mu_j$.  For fixed $p$, $\sigma_{ij}$, if $\mu_j \rightarrow \infty$ and $\mu_i \rightarrow
\infty$, then $Corr(X_i, X_j) \rightarrow$ 1, and if  $\mu_j \rightarrow \infty$ and $\mu_i \rightarrow
- \infty$, then $Corr(X_i, X_j) \rightarrow$ -1.  From (\ref{corr-mgsn}) it also follows that if ${\ve X}$ is a standard $d$-variate
MGSN random variable, i.e. for $i \ne j$, $\mu_i = \mu_j = \sigma_{ij}$ = 0, hence $Corr(X_i, X_j)$ = 0.  Therefore, in this case 
although $X_i$ and $X_j$ are uncorrelated, they are not independent.

Now we would like to compute the multivariate skewness indices of the MGSN distribution.  Different multivariate skewness measures 
have been introduced in the literature.  Among them the skewness index of Mardia \cite{Mardia:1970, Mardia:1974} is the most popular one.  To define 
Mardia's multivariate skewness index let us introduce the following notations of 
a random vector ${\ve X} = (X_1, \ldots, X_d)$.
$$
\mu_{i_1, \ldots, i_s}^{(r_1, \ldots, r_s)} = E \left [ \prod_{k=1}^s (X_{r_k} - \mu_{r_k})^{i_k} \right ],
$$
where $\ds \mu_{r_k} = E(X_{r_k}), k = 1, \ldots, s$.  Mardia \cite{Mardia:1970} defined the multivariate skewness index as
$$
\beta_1 = \sum_{r,s,t = 1}^d \sum_{r',s',t' = 1}^d \sigma^{rr'} \sigma^{ss'} \sigma^{tt'} \ \mu_{111}^{rst} \ \mu_{111}^{r's't'},
$$
here $\sigma^{jk}$ for $j,k = 1, \ldots, d$ denotes the $(j,k)$-th element of the inverse of the dispersion matrix of the random 
vector ${\ve X}$.  In case of MGSN distribution 
\be
\mu_{111}^{lhm} = \frac{1}{p^4} \left \{p(1-p)(2-p) \mu_h \mu_l \mu_m + p^2(1-p)(\mu_m \sigma^{hl} + \mu_l \sigma^{hm} + \mu_h \sigma^{lm})
\right \}.   \label{mgsn-mu}
\ee
It is clear from (\ref{mgsn-mu}) that if $p$ = 1 then $\beta_1$ = 0.  Also if $\mu_j$ = 0 for all $j = 1, \ldots, d$, then 
$\beta_1$ = 0.  Moreover, if $\mu_j \ne 0$ for some $j = 1, \ldots, d$, then the skewness index $\beta_1$ may diverge to 
$\infty$ or $-\infty$ as $p \rightarrow 0$.  Therefore, for MGSN distribution Mardia's multivariate skewness index varies from 
$-\infty$ to $\infty$.

If $\ds \ve X \sim$ MGSN$_d(p, {\ve \mu}, {\ve \Sigma})$ and if 
we denote the mean vector and dispersion matrix of ${\ve X}$, as ${\ve \mu}_{\ve X}$ and 
${\ve \Sigma}_{\ve X}$, respectively, then from (\ref{mgsn-mean}), (\ref{mgsn-var}) and (\ref{mgsn-cov}), we have the following relation:
$$
p \ {\ve \mu}_{\ve X} = {\ve \mu} \ \ \ \  \hbox{and} \ \ \ \ p^2 \ {\ve \Sigma}_{\ve X} = p \ {\ve \Sigma} + (1-p) \ 
{\ve \mu} {\ve \mu}^T.    
$$

The following result provides the canonical correlation between ${\ve X}_1$ and ${\ve X}_2$.  It may be mentioned that canonical 
correlation is very useful in multivariate data analysis.  In an experimental context suppose we take two sets of variables, 
then the canonical correlation can be used to see what is common among these two sets of variables; see for example 
Rao \cite{Rao:1973}.

\noindent {\sc Theorem 2:}  Suppose ${\ve X} \sim$ MGSN$_d(p, {\ve 0}, {\ve \Sigma})$.  Further ${\ve X}$ and ${\ve \Sigma}$ 
are partitioned as in (\ref{partition}).  Then for ${\ve \alpha} \in \mr^h$ and ${\ve \beta} \in \mr^{d-h}$ such that 
${\ve \alpha}^T {\ve \Sigma}_{11} {\ve \alpha} = 1$ and ${\ve \beta}^T {\ve \Sigma}_{22} {\ve \beta} = 1$, the maximum 
corr$({\ve \alpha}^T {\ve X}_1, {\ve \beta}^T {\ve X}_2)$ = $\lambda_1$, where $\lambda_1$ is the maximum root of the $d$-degree 
polynomial equation 
$$
\left | \matrix{- \lambda {\ve \Sigma}_{11} & {\ve \Sigma}_{12} \cr {\ve \Sigma}_{21} & - \lambda {\Sigma}_{22} \cr} \right | = 0.
$$

\noindent {\sc Proof:} From Theorem 1, we obtain
$$
\left ( \matrix{{\ve \alpha}^T {\ve X}_1 \cr {\ve \beta}^T {\ve X}_2 \cr} \right ) \sim \hbox{MGSN} 
\left (2, p, \left ( \matrix{{\ve \alpha}^T {\ve \mu}_1 \cr {\ve \beta}^T {\ve \mu}_2 \cr} \right ), 
\left ( \matrix{{\ve \alpha}^T {\ve \Sigma}_{11} {\ve \alpha} & {\ve \alpha}^T {\ve \Sigma}_{12} {\ve \beta}  \cr
{\ve \beta}^T {\ve \Sigma}_{21} {\ve \alpha} & {\ve \beta}^T {\ve \Sigma}_{22} {\ve \beta} \cr} \right ) \right ).
$$
Therefore, using ({\ref{corr-mgsn}), it follows that the problem is to find ${\ve \alpha} \in \mr^h$ and ${\ve \beta} \in 
\mr^{d-h}$ such that it maximizes
$$
\hbox{corr}({\ve \alpha}^T {\ve X}_1, {\ve \beta}^T {\ve X}_2) = \frac{p{\ve \alpha}^T {\ve \Sigma}_{12} {\ve \beta}^T}
{\sqrt{p{\ve \alpha}^T {\ve \Sigma}_{11} {\ve \alpha}} \sqrt{p{\ve \beta}^T {\ve \Sigma}_{22} {\ve \beta}}} = 
{\ve \alpha}^T {\ve \Sigma}_{12} {\ve \beta}^T,
$$
subject to the restrictions ${\ve \alpha}^T {\ve \Sigma}_{11} {\ve \alpha} = 1$ and ${\ve \beta}^T {\ve \Sigma}_{22} {\ve \beta} 
= 1$.  Now following the same steps as in the multivariate normal cases, Anderson \cite{Anderson:1984}, the result follows.  \qed

The following result provides the characteristic function of the Wishart type matrix based on MGSN random variables.  

\noindent {\sc Theorem 3:} Suppose $Z_1, \ldots, Z_n$ are $n$ i.i.d. random 
variables, and $Z_1 \sim$ MGSN$_d(p,{\ve 0}, {\ve \Sigma})$.  Let us consider the Wishart type matrix
$$
{\ve A} = \sum_{m=1}^n {\ve Z}_m {\ve Z}_m^T = ((A_{ij})), \ \ \ i,j = 1, \ldots, d.
$$
If ${\ve \Theta} = ((\theta_{ij}))$ with $\theta_{ij} = \theta_{ji}$ is a $d \times d$ matrix, then the characteristic function 
of \linebreak $(A_{11}, \ldots, A_{pp}$, $2 A_{12}, 2 A_{13}, \ldots, 2 A_{p-1,p})$ is
$$
E \left ( e^{ i \hbox{tr} \left ({\ve A}{\ve \Theta} \right )} \right ) = p^n 
\left [ \sum_{k=1}^{\infty} \left |{\ve I} - 2 i k {\ve \Theta}{\ve \Sigma} \right |
^{-1/2} (1-p)^{k-1} \right ]^n.
$$

\noindent {\sc Proof:} 
\beanno
E \left [ e^{i \hbox{tr} \left ({\ve A}{\ve \Theta} \right )} \right ] & = & E \left [ e^{i \hbox{tr} \left ( \sum_{m=1}^n 
{\ve Z}_m {\ve Z}_m^T {\ve \Theta} \right )} \right ] = E \left [ e^{i \hbox{tr} \left ( \sum_{m=1}^n 
{\ve Z}_m^T {\ve \Theta} {\ve Z}_m \right )} \right ]  \\
&  = & E \left [ e^{i \left ( \sum_{m=1}^n  {\ve Z}_m^T {\ve \Theta} {\ve Z}_m \right )} \right ] = 
\left ( E \left [ e^{i \left ( {\ve Z}_1^T {\ve \Theta} {\ve Z}_1\right )} \right ] \right )^n.
\eeanno
Now we would like to compute $\ds E \left [ e^{i \left ( {\ve Z}_1^T {\ve \Theta} {\ve Z}_1\right )} \right ]$.  For a $d \times d$ 
real symmetric matrix ${\ve \Theta}$, there is a real $d \times d$ matrix ${\ve B}$ such that
$$
{\ve B}^T {\ve \Sigma}^{-1} {\ve B} = {\ve I} \ \ \ \ \hbox{and} \ \ \ \ {\ve B}^T {\ve \Theta} {\ve B} = {\ve D} = 
\hbox{diag} \{\delta_1, \ldots, \delta_d\}.
$$
Here, $\ds \hbox{diag} \{\delta_1, \ldots, \delta_d\}$ means a $d \times d$ diagonal matrix with diagonal entries as 
$\delta_1, \ldots, \delta_d$.  If we make the transformation ${\ve Z}_1 = {\ve B} {\ve Y}$, then using Theorem 1, 
${\ve Y} \sim$ MGSN$_d(p, {\ve 0}, {\ve I})$.  
Using the definition MGSN distribution it follows that 
$$
{\ve Y} \stackrel{d}{=} \sum_{m=1}^N {\ve Y}_m,
$$
here $N \sim$ GE$(p)$, and ${\ve Y}_m$'s are i.i.d. random vectors, and ${\ve Y}_1 \sim$ N$_d({\ve 0}, {\ve I})$.  We use the 
following notation
$$
{\ve Y} = (Y_1, \ldots, Y_d)^T \ \ \  \hbox{and} \ \ \  {\ve Y}_m = (Y_{m1}, \ldots, Y_{md})^T.
$$
Hence, $\ds Y_j = \sum_{m=1}^N Y_{mj}$, for $j = 1, \ldots, d$.
Therefore,
\beanno
E \left [ e^{i \left ( {\ve Z}_1^T {\ve \Theta} {\ve Z}_1\right )} \right ] & = &  
E \left [ e^{i \left ( {\ve Y}^T {\ve D} {\ve Y} \right )} \right ] = E \left [ e^{i \left ( \sum_{j=1}^d \delta_j Y_j^2 \right )} \right ] =
E \left [ e^{i \left ( \sum_{j=1}^d \delta_j \left ( \sum_{m=1}^N Y_{mj} \right )^2 \right )} \right ]  \\
& = & E_N E \left [ \left . e^{i \left ( \sum_{j=1}^d \delta_j \left ( \sum_{m=1}^N Y_{mj} \right )^2 \right )} \right | N \right ]  
 =  E_N E \left [ \left . e^{i \left ( \sum_{j=1}^d \delta_j N \left ( \sum_{m=1}^N Y_{mj} /\sqrt{N} \right )^2 \right )} \right | N \right ]  \\
& = & E_N \prod_{j=1}^d E \left [ \left . e^{i \left (\delta_j N \left ( \sum_{m=1}^N Y_{mj} /\sqrt{N} \right )^2 \right )} \right | N \right ]  
 =  E_N \prod_{j=1}^d \left ( 1 - 2 i \delta_j N \right )^{-1/2}  \\
& = & E_N \left | {\ve I} - 2 i N {\ve D}  \right |^{-1/2}  
 =  E_N \left | {\ve I} - 2 i N {\ve \Theta} {\ve \Sigma}  \right |^{-1/2}  \\
& = & p \sum_{k=1}^{\infty} \left |{\ve I} - 2 i k {\ve \Theta}{\ve \Sigma} \right |
^{-1/2} (1-p)^{k-1}.
\eeanno
\qed

\noindent {\sc Theorem 4:} If for any ${\ve c} \ne {\ve 0}, {\ve c} \in \mr^d$, ${\ve c}^T {\ve X} \sim$ 
GSN$(\mu({\ve c}), \sigma({\ve c}), p)$ for a $d$ dimensional random vector ${\ve X}$, then there exists  a $d$-dimensional 
vector ${\ve \mu}$ and a $d \times d$ symmetric matrix ${\ve \Sigma}$ such that  ${\ve X} \sim$ MGSN$_d(p, {\ve \mu}, {\ve \Sigma})$.  Here $ -\infty < \mu({\ve c}) < \infty, 0 < \sigma({\ve c}) < \infty$ are functions of ${\ve c}$.

\noindent {\sc Proof:} If we denote the mean vector and dispersion matrix of the random vector ${\ve X}$, as ${\ve \mu}_{\ve X}$ and 
${\ve \Sigma}_{\ve X}$, respectively, then we have $E({\ve c}^T {\ve X}) = {\ve c}^T {\ve \mu}_{\small{\ve X}}$ and $V({\ve c}^T {\ve X}) = {\ve c}^T {\ve \Sigma}_{\ve X} {\ve c}$.  Hence from (\ref{mean-gsn}) and (\ref{var-gsn}), we have 
\be
\mu({\ve c}) = p {\ve c}^T {\ve \mu}_{\ve X} \ \ \  \hbox{and} \ \ \ 
\sigma^2({\ve c}) = p {\ve c}^T {\ve \Sigma}_{\ve X} {\ve c} - p(1-p) \left ( {\ve c}^T {\ve \mu}_{\ve X} \right )^2.
  \label{rel-2}
\ee Therefore, 
from (\ref{mgf-gsn}), using $t$ = 1, it follows that
\be
E \left ( \exp ({\ve c}^T {\ve X}) \right ) = \frac{p \exp (\mu({\ve c}) + \sigma^2({\ve c})/2)}{1 - (1-p)\exp (\mu({\ve c}) + \sigma^2({\ve c})/2)}.      \label{mgfc}
\ee
Let us define a $d$-dimensional vector {\ve \mu} and a $d \times d$ symmetric  matrix ${\ve \Sigma}$ as given below:
\be
{\ve \mu} = p {\ve \mu}_{\ve X} \ \ \ \hbox{and} \ \ \ \ 
{\ve \Sigma} = p {\ve \Sigma}_{\ve X} - p(1-p) {\ve \mu}_{\ve X} {\ve \mu}_{\ve X}^T.
\label{rel-1}
\ee 
Therefore, 
$$
\mu({\ve c}) + \frac{ \sigma^2({\ve c}) }{2} = {\ve c}^T {\ve \mu} + \frac{1}{2} {\ve c}^T {\ve \Sigma} {\ve c},
$$
and (\ref{mgfc}) can be written as
$$
E \left ( e^{{\ve c}^T {\ve X}} \right ) = \frac{p \exp \left ({\ve c}^T {\ve \mu} + \frac{1}{2} {\ve c}^T {\ve \Sigma} {\ve c} 
\right )}{1 - (1-p) 
\exp \left ({\ve c}^T {\ve \mu} + \frac{1}{2} {\ve c}^T {\ve \Sigma} {\ve c} \right )} = M_{\ve X}({\ve c}).
$$
Hence the result follows.   \qed

Therefore, combining Theorems 1 and 2, we have the following characterization results for a $d$-variate MGSN distribution.

\noindent {\sc Theorem 5:} If a $d$-dimensional random vector ${\ve X}$ has a mean vector ${\mu}_{\ve X}$ and a dispersion 
matrix ${\ve \Sigma}_{\ve X}$, then ${\ve X} \sim$ MGSN$_d(p, {\ve \mu}, {\ve \Sigma})$, here ${\ve \mu}$ and ${\ve \Sigma}$ are as 
defined in (\ref{rel-1}), if and only if for any ${\ve c} \ne {\ve 0}, {\ve c} \in \mr^d$, ${\ve c}^T {\ve X} \sim$ 
GSN$(\mu({\ve c}), \sigma({\ve c}), p)$, where $\mu({\ve c})$ and $\sigma({\ve c})$ are as in  (\ref{rel-2}).

Now we provide another characterization of the MGSN distribution.  

\noindent {\sc Theorem 6:} Suppose ${\ve X}_1, {\ve X}_2, \ldots$ is a sequence of i.i.d. $d$-dimensional random vectors, 
and $M \sim$ GE$(\alpha)$, for $0 <\alpha \le 1$.  Consider a new $d$-dimensional random vector
$$
{\ve Y} = \sum_{i=1}^M {\ve X}_i.
$$
Then ${\ve Y} \sim$ MGSN$_d(\beta, {\ve \mu}, {\ve \Sigma})$ for $\beta \le \alpha$, if and only if ${\ve X}_1$ has a MGSN distribution.

\noindent {\sc Proof:} If part. Suppose ${\ve X}_1 \sim$ MGSN$_d(p,{\ve \mu}, \ve{\Sigma})$, then the MGF of ${\ve Y}$ for 
${\ve t} \in \mr^d$, can be written as
\beanno
M_{\ve Y}({\ve t}) & = & E \left (e^{{\ve t}^T {\ve Y}} \right ) = \sum_{m=1}^{\infty} E  \left (e^{\sum_{i=1}^M {\ve t}^T  {\ve X}_i}| M = m \right )P(M=m)  \\
& = & \sum_{m=1}^{\infty} \alpha (1-\alpha)^{m-1} \left (  \frac{p e^{{\ve \mu}^T {\ve t} + \frac{1}{2} {\ve t}^T {\ve \Sigma} {\ve t}}}{1 - (1-p)  e^{{\ve \mu}^T {\ve t} + \frac{1}{2} {\ve t}^T {\ve \Sigma} {\ve t}}} \right )^m  \\
& = &  \frac{\alpha p e^{{\ve \mu}^T {\ve t} + \frac{1}{2} {\ve t}^T {\ve \Sigma} {\ve t}}}{1 - (1-\alpha p)  e^{{\ve \mu}^T {\ve t} + \frac{1}{2} {\ve t}^T {\ve \Sigma} {\ve t}}} 
 =   \frac{\beta e^{{\ve \mu}^T {\ve t} + \frac{1}{2} {\ve t}^T {\ve \Sigma} {\ve t}}}{1 - (1- \beta)  e^{{\ve \mu}^T {\ve t} + \frac{1}{2} {\ve t}^T {\ve \Sigma} {\ve t}}},
\eeanno
here $\beta = \alpha p \le \alpha$.

Only if part.  Suppose ${\ve Y} \sim$ MGSN$_d(\beta, {\ve \mu}, {\ve \Sigma})$ for some $0 < \beta \le \alpha$, and the MGF 
of ${\ve X}_1$ is $M_{{\ve X}_1}({\ve t})$.  We have the following relation:
\be
\frac{\beta e^{{\ve \mu}^T {\ve t} + \frac{1}{2} {\ve t}^T {\ve \Sigma} {\ve t}}}{1 - (1-\beta)  e^{{\ve \mu}^T {\ve t} + \frac{1}{2} {\ve t}^T {\ve \Sigma} {\ve t}}}
= \frac{\alpha M_{{\ve X}_1}({\ve t})}{1 - (1-\alpha)M_{{\ve X}_1}({\ve t})}.    \label{two-mgf}
\ee
From (\ref{two-mgf}), we obtain
$$
M_{{\ve X}_1}({\ve t}) = \frac{\gamma e^{{\ve \mu}^T {\ve t} + \frac{1}{2} {\ve t}^T {\ve \Sigma} {\ve t}}}{1 - (1-\gamma)  e^{{\ve \mu}^T {\ve t} + \frac{1}{2} {\ve t}^T {\ve \Sigma} {\ve t}}},
$$
for $\gamma = \beta/\alpha \le 1$.  Therefore,
${\ve X}_1 \sim$ MGSN$_d(\gamma,{\ve \mu}, {\ve \Sigma})$.   \qed

  Stochastic ordering plays a very important role in the distribution 
theory.  It has been studied quite extensively in the statistical literature.  
For its importance and different applications, 
interested readers are referred to Shaked and Shantikumar \cite{SS:1994}.
Now we will discuss the multivariate total positivity of order two (MTP$_2$) property, in the sense of Karlin and Rinott 
\cite{KR:1980}, of the joint PDF of MGSN distribution.
  We shall be using the following notation here.  For any two real numbers
$a$ and $b$, let $a \wedge b$ = min$\{a,b\}$, and $a \vee b$ = max$\{a,b\}$.  For any vector ${\ve x} = (x_1, \ldots, x_d)^T$ 
and ${\ve y} = (y_1, \ldots, y_d)^T$, let ${\ve x} \vee {\ve y} = (x_1 \vee y_1, \ldots, x_d \vee y_d)^T$ and 
${\ve x} \wedge {\ve y} = (x_1 \wedge y_1, \ldots, x_d \wedge y_d)^T$.  Let us recall the definition of MTP$_2$ 
property.  A function
$g: \mr^d \rightarrow \mr^+$ is said to have MTP$_2$ property, in the sense of Karlin and Rinott \cite{KR:1980}, if 
$\ds g({\ve x}) g({\ve y}) \le g({\ve x} \wedge {\ve y}) g({\ve x} \vee {\ve y})$, for all ${\ve x}, {\ve y} \in \mr^d$.  We then 
have the following result for MGSN distribution.

\noindent {\sc Theorem 7:} Let ${\ve X} \sim$ MGSN$_d(p, {\ve 0}, {\ve \Sigma})$, and all the off-diagonal elements of 
${\ve \Sigma}^{-1}$ are less than or equal to zero, then the PDF of ${\ve X}$ has MTP$_2$ property.

\noindent {\sc Proof:} To prove that the PDF of ${\ve X}$ has MTP$_2$ property, it is enough to show that
\be
{\ve x}^T {\ve \Sigma}^{-1} {\ve x} + {\ve y}^T {\ve \Sigma}^{-1} {\ve y} \ge 
({\ve x} \vee {\ve y})^T {\ve \Sigma}^{-1} ({\ve x} \vee {\ve y}) + ({\ve x} \wedge {\ve y})^T {\ve \Sigma}^{-1} ({\ve x} 
\wedge {\ve y})   \label{tp2}
\ee
for any ${\ve x} = (x_1, \ldots, x_d)^T \in \mr^d$ and ${\ve y} = (y_1, \ldots, y_d)^T \in \mr^d$.  If the elements of ${\ve 
\Sigma}^{-1}$ are denoted by $((\sigma^{kj}))$, for $k,j = 1, \ldots, d$, then proving (\ref{tp2}) is equivalent to showing
$$
\sum_{\stackrel{k,j=1}{k \ne j}}^d (x_j x_k + y_j y_k)\sigma^{jk} \ge \sum_{\stackrel{k,j=1}{k \ne j}}^d ((x_j \wedge y_j)(x_k \wedge y_k) 
+ (x_j \vee y_j)(x_k \vee  y_k))\sigma^{jk}.
$$
For all $k,j = 1, \ldots, d$,
$$
(x_j x_k + y_j y_k) \le (x_j \wedge y_j)(x_k \wedge y_k) + (x_j \vee y_j)(x_k \vee  y_k),
$$
which can be easily shown by taking any ordering of $x_k, x_j, y_k, y_j$.  Now the result follows since $\sigma^{jk} \le 0$.
\qed

The following two decompositions of a MGSN distribution are possible.  We use the following notations.  The distribution of a 
negative binomial random 
variable with parameters $r$ and $p$, where $r$ is a non-negative integer and $0 < p < 1$, will be denoted by NB$(r,p)$.  
If $T \sim$ NB$(r,p)$, then the MGF of $T$ is
$$
M_T(t) = \left ( \frac{1-p}{1-pe^t} \right )^r \ \ \ \hbox{for} \ \ \ t < - \ln p.
$$
A discrete random variable $Z$ is said to have a logarithmic distribution with parameter $p$, for $0 < p < 1$, if the 
PMF of $Z$ is 
$$
P(Z = k) = \frac{(1-p)^k}{\lambda k}  \ \ \ \ \hbox{for} \ \ \ k = 1, 2, \ldots, \ \ \ \hbox{where} \ \ \  \lambda = -\ln p,
$$
and it will be denoted by LD$(p)$.  Now we provide two decompositions of MGSN distribution.

\noindent {\sc Decomposition 1:}  Suppose ${\ve X} \sim$ MGSN$_d(p,{\ve \mu}, {\ve \Sigma})$.  Further, for  
any positive integer $n$ and for any $1 \le k \le n$, suppose
$$
{\ve Z}_{kn} \stackrel{disp}{=} \sum_{j=1}^{1+nT} {\ve Y}_j,
$$
where $T \sim$ NB$(r,p)$ and $r = 1/n$, ${\ve Y}_j$'s are i.i.d. N$_d(r{\ve \mu}, r {\ve \Sigma})$, $T$ and ${\ve Y}_j$'s are
independently distributed, then
$$
{\ve X} \stackrel{disp}{=} {\ve Z}_{1n} + \ldots + {\ve Z}_{nn}.
$$

\noindent {\sc Proof:} The MGF of ${\ve Z}_{kn}$ can be written as
\beanno
M_{{\ve Z}_{kn}}({\ve t}) & =  & E \left ( e^{{\ve t}^T {\ve Z}_{kn}} \right ) = \sum_{j=0}^{\infty} E \left ( e^{{\ve t}^T {\ve Z}_{kn}} |T = j \right )
P(T = j)  \\
& = & \left [ \frac{p e^{{\ve \mu}^T {\ve t} + \frac{1}{2} {\ve t}^T {\ve \Sigma} {\ve t}}}{1 - (1-p) e^{{\ve \mu}^T {\ve t} + \frac{1}{2} {\ve t}^T {\ve \Sigma} {\ve t}}}
\right ]^r = \left [ M_{\ve X}({\ve t}) \right ]^r.    
\eeanno
\qed

\noindent It implies that MGSN law is infinitely divisible.  The following decomposition is also possible.

\noindent {\sc Decomposition 2:} Suppose   $Q$ is a Poisson random variable with 
parameter $\lambda$, and \linebreak $\{Z_i, i = 1, 2, \ldots\}$ is a sequence of i.i.d. random variables having logarithmic distribution 
with the following probability mass function for $\lambda = -\ln p$;
$$
P(Z_1 = k) = \frac{(1-p)^k}{\lambda k}; \ \ \ k = 1, 2, \ldots,
$$
and all the random variables are independently distributed.  If ${\ve X} \sim$ MGSN$_d(p,{\ve \mu}, {\ve \Sigma})$,
then the following decomposition is possible
\be
{\ve X} \stackrel{disp}{=} {\ve Y} + \sum_{i=1}^Q {\ve Y}_i,    \label{decom-2}
\ee
here $\{{\ve Y}_i| Z_i = k\} \sim$ N$_d(k {\ve \mu}, k {\ve \Sigma})$ for $i = 1, 2, \ldots$, and they are independently distributed, 
${\ve Y} \sim$ N$_d({\ve \mu}, {\ve \Sigma})$, and it is independent of $Q$, and $({\ve Y}_i, Z_i)$ for all $i = 1, 2, \ldots.$

\noindent {\sc Proof:} First note that the probability generating function of $Q$ and $Z_1$ are as follows:
$$
E(t^Q) = e^{\lambda(t-1)} \ \ \ \ \hbox{and} \ \ \ \ E(t^{Z_1}) = \frac{\ln(1 - (1-p)t)}{\ln p}; \ \ \  t < (1-p)^{-1}.
$$
The MGF of ${\ve Y}_i$ for ${\ve t} \in \mr^d$, such that $\ds (1-p) e^{{\ve \mu}^T {\ve t} + \frac{1}{2} {\ve t}^T {\ve \Sigma} {\ve t}} < 1$,
can be obtained as
$$
M_{{\ve Y}_i}({\ve t}) = E \left ( e^{{\ve t}^T {\ve Y}_i} \right ) = E_{Z_i} E_{{\ve Y}_i|Z_i} \left ( e^{{\ve t}^T {\ve Y}_i} \right )
= \frac{\ln \left (1 - (1-p) e^{{\ve \mu}^T {\ve t} + \frac{1}{2} {\ve t}^T {\ve \Sigma} {\ve t}} \right )}{\ln p}.
$$
Therefore, the MGF of the right hand side of (\ref{decom-2}) can be written as
\beanno
E \left [ e^{{\ve t}^T \left ({\ve Y} + \sum_{i=1}^Q {\ve Y}_i \right )} \right ] & = & 
e^{{\ve \mu}^T {\ve t} + \frac{1}{2} {\ve t}^T {\ve \Sigma} {\ve t}} \times 
E \left [ 
\frac{\ln \left (1 - (1-p) e^{{\ve \mu}^T {\ve t} + \frac{1}{2} {\ve t}^T {\ve \Sigma} {\ve t}} \right )}{\ln p} 
\right ]^Q \\
& = & \frac{p e^{{\ve \mu}^T {\ve t} + \frac{1}{2} {\ve t}^T {\ve \Sigma} {\ve t}}}{1 - (1-p)e^{{\ve \mu}^T {\ve t} + \frac{1}{2} {\ve t}^T {\ve \Sigma} {\ve t}}}
= M_{\ve X}({\ve t}).
\eeanno
\qed

The following results will be useful for further development.  Let us consider the random vector $({\ve X}, N)$, where ${\ve X}$
and $N$ are same as defined in (\ref{mgsn-def}).  The joint PDF of $({\ve X}, N)$ can be written as
$$
f_{{\ve X}, N}({\ve x}, n) = \left \{ \matrix{
\frac{p(1-p)^{n-1}}{(2 \pi)^{d/2} \left |{\ve \Sigma} \right |^{1/2} n^{d/2}}
e^{-\frac{1}{2n} ({\ve x} - n {\ve \mu})^T {\ve \Sigma}^{-1}({\ve x} -n {\ve \mu})} & \hbox{if} & 0 < p < 1 \cr
&  &  \cr
\frac{1}{(2 \pi)^{d/2} \left |{\ve \Sigma} \right |^{1/2}}
e^{-\frac{1}{2} ({\ve x} - {\ve \mu})^T {\ve \Sigma}^{-1}({\ve x} - {\ve \mu})} & \hbox{if} & p = 1, \cr}  \right .
$$
for ${\ve x} \in \mr^d$ and for any positive integer $n$.  Therefore, the conditional probability mass function of $N$ given 
${\ve X} = {\ve x}$ becomes
$$
P(N = n|{\ve X} = {\ve x}) = \frac{ (1-p)^{n-1} e^{-\frac{1}{2n} ({\ve x} - n {\ve \mu})^T {\ve \Sigma}^{-1}({\ve x} -n {\ve \mu})} n^{-d/2}
}{\sum_{k=1}^{\infty} (1-p)^{k-1}  e^{-\frac{1}{2k} ({\ve x} - k {\ve \mu})^T {\ve \Sigma}^{-1}({\ve x} -k {\ve \mu})} k^{-d/2}}.
$$
Therefore,
\be
E(N|{\ve X} = {\ve x}) = \frac{\sum_{n=1}^{\infty}(1-p)^{n-1}   e^{-\frac{1}{2n} ({\ve x} - n {\ve \mu})^T {\ve \Sigma}^{-1}({\ve x} -n {\ve \mu})}n^{-d/2+1}}
{\sum_{k=1}^{\infty} (1-p)^{k-1}  e^{-\frac{1}{2k} ({\ve x} - k {\ve \mu})^T {\ve \Sigma}^{-1}({\ve x} -k {\ve \mu})} k^{-d/2}},    \label{expn}
\ee
and
\be
E(N^{-1}|{\ve X} = {\ve x}) = \frac{\sum_{n=1}^{\infty}(1-p)^{n-1} e^{-\frac{1}{2n} ({\ve x} - n {\ve \mu})^T {\ve \Sigma}^{-1}({\ve x} -n {\ve \mu})}n^{-d/2-1}}
{\sum_{k=1}^{\infty} (1-p)^{k-1}  e^{-\frac{1}{2k} ({\ve x} - k {\ve \mu})^T {\ve \Sigma}^{-1}({\ve x} -k {\ve \mu})} k^{-d/2}}.   \label{expnsq}
\ee

\section{\sc Statistical Inference}

\subsection{\sc Estimation}

In this section we discuss the maximum likelihood estimators (MLEs) of the unknown parameters, when $0 < p < 1$.  When $p$ = 1, the 
MLEs of ${\ve \mu}$ and ${\ve \Sigma}$ can be easily obtained as the sample mean and the sample variance covariance matrix, respectively.
Suppose ${\cal D} = \{{\ve x}_1, \ldots, {\ve x}_n\}$ is a random sample of size $n$ from MGSN$_d(p, {\ve \mu}, {\ve \Sigma})$, then the
log-likelihood function becomes
\bea
l(p, {\ve \mu}, {\ve \Sigma}) & = & \sum_{i=1}^n \ln f_{\ve X}({\ve x}_i; {\ve \mu}, {\ve \Sigma}, p) \nonumber  \\
& = & \sum_{i=1}^n \ln \left [\sum_{k=1}^{\infty} \frac{p(1-p)^{k-1}}{(2 \pi)^{d/2} |{\ve \Sigma}|^{1/2} k^{d/2}}
e^{-\frac{1}{2k} ({\ve x}_i - k {\ve \mu})^T {\ve \Sigma}^{-1}({\ve x}_i - k {\ve \mu})} \right ].  \label{ll-mgsn}
\eea
The maximum likelihood estimators (MLEs) of the unknown parameters can be obtained by maximizing (\ref{ll-mgsn}) with respect 
to the unknown parameters.  It involves solving a $(d+1+d(d+1)/2)$ dimensional optimization problem.  Therefore, for large $d$, it is a challenging issue.

To avoid that problem, first it is assumed that $p$ is known.  For a known $p$, we estimate the MLEs of ${\ve \mu}$ and 
${\ve \Sigma}$ by using EM algorithm, say $\widehat{\ve \mu}(p)$ and $\widehat{\ve \Sigma}(p)$, respectively.  We maximize 
$l(p, \widehat{\ve \mu}(p), \widehat{\ve \Sigma}(p))$ to compute $\widehat{p}$, the MLE of $p$.  Finally we obtain the MLE of 
${\ve \mu}$ and ${\ve \Sigma}$ as $\widehat{\ve \mu} = \widehat{\ve \mu}(\widehat{p})$ and $\widehat{\ve \Sigma} = 
\widehat{\ve \Sigma}(\widehat{p})$, respectively.  Now we will show how to compute $\widehat{\ve \mu}(p)$ and $\widehat{\ve \Sigma}(p)$,
for a given $p$ using EM algorithm.  We treat the problem as a missing value problem, and the main idea is as follows.

It is assumed that $p$ is known.  Suppose we have the complete observations of the form $\{({\ve x}_1, m_1), \ldots, ({\ve x}_n, m_n)\}$ 
from $({\ve X}, N)$.  Then the log-likelihood function based on the complete observation becomes (without the additive constant)
$$
l_c({\ve \mu}, {\ve \Sigma})  =  - \frac{n}{2} \ln |{\ve \Sigma}| 
- \frac{1}{2} \sum_{i=1}^n \frac{1}{m_i} ({\ve x}_i - m_i {\ve \mu})^T {\ve \Sigma}^{-1} ({\ve x}_i - m_i {\ve \mu}).   
$$
Therefore, if we define the MLEs of ${\ve \mu}$ and ${\ve \Sigma}$ based on the complete observations as  
$\widehat{\ve \mu}_c(p)$ and $\widehat{\Sigma}_c(p)$, respectively, then for $\displaystyle K = \sum_{i=1}^n m_i$,
\be
\widehat{\ve \mu}_c(p) = \frac{1}{K} \sum_{i=1}^n {\ve x}_i,   \label{mle1-c}
\ee
and
\bea
\widehat{\ve \Sigma}_c(p) & = & \frac{1}{n} \sum_{i=1}^n  \frac{1}{m_i} ({\ve x}_i - m_i \widehat{\ve \mu}_c) ({\ve x}_i - m_i \widehat{\ve \mu}_c)^T 
\nonumber  \\ & = &
\frac{1}{n} \left [ \sum_{i=1}^n \frac{1}{m_i} {\ve x}_i {\ve x}^T_i -\sum_{i=1}^n ( \widehat{\ve \mu}_c   {\ve x}^T_i + {\ve x}_i 
\widehat{\ve \mu}_c^T) 
 + K \widehat{\ve \mu}_c \widehat{\ve \mu}^T_c  
\right ].        \label{mle2-c}
\eea
Note that $\displaystyle \widehat{\ve \mu}_c(p)$ is obtained by taking 
derivative of $\displaystyle \sum_{i=1}^n \frac{1}{m_i} ({\ve x}_i - m_i {\ve \mu})^T {\ve \Sigma}^{-1} ({\ve x}_i - m_i {\ve \mu})$ with respect to 
${\ve \mu}$, and equate it to zero.  Similarly, $\displaystyle \widehat{\ve \Sigma}_c(p)$ is obtained by using Lemma 3.2.2 of Anderson 
\cite{Anderson:1984}.

Now we are ready to provide the EM algorithm for a given $p$.  The EM algorithm consists of maximizing the conditional expectation of the complete 
log-likelihood function, based on the observed data and the current value of ${\ve \theta} = ({\ve \mu}, {\ve \Sigma})$, say 
$\widetilde{\ve \theta}$, in an iterative two-step algorithm process, see for example Dempster et al. \cite{DLR:1977}.   The E-step is to compute
the conditional expectation denoted by $Q({\ve \theta}|\widetilde{\ve \theta})$, and the M-step is maximizing $Q({\ve \theta}|\widetilde{\ve \theta})$, with respect to ${\ve \theta}$.  We use the following notations:
$$
a_i = E(N|{\ve X} = {\ve x}_i, \widetilde{\ve \theta}) \ \ \ \ \ \hbox{and} \ \ \ \ b_i = E(N^{-1}|{\ve X} = {\ve x}_i, \widetilde{\ve \theta}),
$$
where $a_i$ and $b_i$ are obtained using (\ref{expn}) and (\ref{expnsq}), respectively.

\noindent {\sc E-Step:} It consists of calculating $Q({\ve \theta}|\widetilde{\ve \theta})$, $\widetilde{\ve \theta}$ being the current 
parameter value.
\beanno
Q({\ve \theta}|\widetilde{\ve \theta})  & = & E(l_c({\ve \theta}|{\cal D}, \widetilde{\ve \theta}))  \nonumber \\  & = & 
- \frac{n}{2} \ln |{\ve \Sigma}|   - \frac{1}{2} \hbox{trace} \left \{ {\ve \Sigma}^{-1} \left (  \sum_{i=1}^n b_i {\ve x}_i {\ve x}^T_i - 
 \sum_{i=1}^n ( {\ve x}_i {\ve \mu}^T + {\ve \mu} {\ve x}_i^T) + 
{\ve \mu} {\ve \mu}^T  \sum_{i=1}^n a_i \right ) \right \}.    \label{e-step}
\eeanno

\noindent {\sc M-Step:} It involves maximizing $Q({\ve \theta}|\widetilde{\ve \theta})$ with respect to ${\ve \theta}$, to obtain 
$\overline{\ve \theta}$, where 
$$
\overline{\ve \theta} = \hbox{arg max}_{\ve \theta} Q({\ve \theta}|\widetilde{\ve \theta}).
$$ 
Here, $\ds \hbox{arg max}_{\ve \theta} Q({\ve \theta}|\widetilde{\ve \theta})$ means the value of ${\ve \theta}$ for which the function
$\ds Q({\ve \theta}|\widetilde{\ve \theta})$ takes the maximum value.
From (\ref{mle1-c}) and (\ref{mle2-c}), we obtain
$$
\overline{\ve \mu} = \frac{1}{\sum_{j=1}^n a_j} \sum_{i=1}^n {\ve x}_i    
$$
and
\be
\overline{\ve \Sigma}  =  
\frac{1}{n} \left [ \sum_{i=1}^n b_i {\ve x}_i {\ve x}^T_i - \sum_{i=1}^n ( {\ve x}_i  \overline{\ve \mu}^T  + 
 \overline{\ve \mu} {\ve x}_i^T) +  
\overline{\ve \mu} \ \overline{\ve \mu}^T \sum_{i=1}^n a_i   
\right ].     \label{m-step-2}
\ee
We propose the following algorithm to compute the MLEs of ${\ve \mu}$ and ${\ve \Sigma}$ for a known $p$.

\noindent {\sc Algorithm 2:}
\begin{itemize}
\item Step 1: Choose an initial guess of ${\ve \theta}$, say ${\ve \theta}^{(0)}$.
\item Step 2: Obtain
$$
{\ve \theta}^{(1)} = \hbox{arg max}_{\ve \theta} Q({\ve \theta}|{\ve \theta}^{(0)}).
$$

\item Step 3: Continue the process until convergence takes place.
\end{itemize}

Once for a given $p$, the MLEs of ${\ve \mu}$ and ${\ve \Sigma}$ are obtained, say $\widehat{\ve \mu}(p)$ and $\widehat{\ve \Sigma}(p)$, 
respectively, then the MLE of $p$ can be obtained by maximizing the profile log-likelihood function of $p$, i.e. 
$\ds l(p,\widehat{\ve \mu}(p), \widehat{\ve \Sigma}(p))$, with respect to $p$.  If it is denoted by $\widehat{p}$, then the MLEs of 
${\ve \mu}$ and ${\ve \Sigma}$ become $\widehat{\ve \mu} = \widehat{\ve \mu}(\widehat{p})$ and 
$\widehat{\ve \Sigma} = \widehat{\ve \Sigma}(\widehat{p})$, respectively.  The details will be explained in Section \ref{sda}.
We have used
the sample mean vector and the sample variance covariance matrix as the initial guess of ${\ve \mu}$ and ${\ve \Sigma}$, 
respectively, of the proposed EM algorithm for all $p$.

\subsection{\sc Testing of Hypotheses}

In this section we discuss three different testing of hypotheses problems which can be useful in practice.  We propose to use the 
likelihood ratio test (LRT) in all the cases, and we indicate the asymptotic distribution of the LRT tests under the 
null hypothesis in each case.  With the abuse of notations, in each case if $\delta$ is any unknown parameter, the MLE of $\delta$
under the null hypothesis will be denoted by $\widehat{\delta}_H$.

\noindent {\sc Test 1:}
\be
H_0: p = 1 \ \ \ \hbox{vs.} \ \ \ \ H_1: p < 1.    \label{test-1}
\ee
The above testing problem (\ref{test-1}) is important in practice as it tests the normality of the distribution.  In this
$\widehat{\ve \mu}_H$ and $\widehat{\ve \Sigma}_H$, respectively,  can be obtained as the sample mean and the sample
variance covariance matrix.  Since in this case $p$ is in the boundary, the standard results do not work.  But using result 3
of Self and Liang \cite{SL:1987} it follows that under the null hypothesis
$$
T_1 = 2(l(\widehat{p}, \widehat{\ve \mu}, \widehat{\ve \Sigma}) - l(1, \widehat{\ve \mu}_H, \widehat{\ve \Sigma}_H)) \longrightarrow
\frac{1}{2} + \frac{1}{2} \chi_1^2.     
$$

\noindent {\sc Test 2:} 
\be
H_0: {\ve \mu} = {\ve 0} \ \ \ \hbox{vs.} \ \ \ \ H_1: {\ve \mu} \ne {\ve 0}.   \label{test-2}
\ee
The above testing problem (\ref{test-2}) is important as it tests the symmetry of the distribution.  In this case under the null
hypothesis the MLEs of $p$ and ${\ve \Sigma}$ can be obtained as follows.  For a given $p$, the MLE of ${\ve \Sigma}$ can be 
obtained using the EM algorithm as before, and then the MLE of $p$ can be obtained by maximizing the profile likelihood function.
In this case the 'E-step' and 'M-Step' can be obtained from (\ref{e-step}) and (\ref{m-step-2}), respectively, 
by replacing ${\ve \mu} = {\ve 0}$.  Under $H_0$, then 
$$
T_2 = 2(l(\widehat{p}, \widehat{\ve \mu}, \widehat{\ve \Sigma}) - l(\widehat{p}_H, {\ve 0}, \widehat{\Sigma}_H)) \longrightarrow
\chi_d^2.      
$$

\noindent {\sc Test 3:}
\be
H_0: {\ve \Sigma} \hbox{   is a diagonal matrix} \ \ \ \ \hbox{vs.} \ \ \ \ {\ve \Sigma} \hbox{   is arbitrary}.   \label{test-3}
\ee
The above testing problem (\ref{test-3}) is important as it tests the uncorrelatedness of the components.  In this case the diagonal 
elements of the matrix ${\ve \Sigma}$ will be denoted by $\sigma_1^2, \ldots, \sigma_d^2$, i.e.
$\ds {\ve \Sigma} = \hbox{diag} \left \{\sigma_1^2, \ldots, \sigma_d^2 \right \}$.
Now we will mention how to compute the MLEs of the unknown parameters $p$, ${\ve \mu}$ and $\sigma_1^2, \ldots, \sigma_d^2$, under
the null hypothesis.  In this case also as before for a given $p$, we use EM algorithm to compute the MLEs of 
${\ve \mu}$ and $\sigma_1^2, \ldots, \sigma_d^2$, and finally the MLE of $p$ can be obtained by maximizing the profile likelihood 
function.  Now we will describe how to compute the MLEs of ${\ve \mu}$ and $\sigma_1^2, \ldots, \sigma_d^2$, for a given $p$, 
by using the EM algorithm.  We use the following notation for further development.  The matrix $\ds {\ve \Delta}_k$ is a 
$d \times d$ matrix with all the entries 0, except the $(k,k)$-th element which is 1.  Now under $H_0$, the `E-Step' of the EM 
algorithm can be written as follows:
\bea
Q({\ve \theta}|\widetilde{\ve \theta}) & = & -\frac{n}{2} \left (\sum_{k=1}^d \ln \sigma_k^2 \right )
\nonumber \\
&  & 
- \frac{1}{2} \sum_{k=1}^d \frac{1}{\sigma_k^2} \left \{
 \sum_{i=1}^n  b_i {\ve x}_i^T {\ve \Delta}_k {\ve x}_i
- \sum_{i=1}^n  \left ( {\ve x}_i^T {\ve \Delta}_k {\ve \mu} + 
 {\ve \mu}^T {\ve \Delta}_k {\ve x}_i \right ) + {\ve \mu}^T {\ve \Delta}_k {\ve \mu} \sum_{i=1}^n a_i 
\right \}.    \label{e-step-3}
\eea
The `M-Step' involves maximizing (\ref{e-step-3}) with respect to ${\ve \mu}$, $\sigma_1^2, \ldots, \sigma_d^2$ to 
obtain updated ${\ve \mu}$, $\sigma_1^2, \ldots, \sigma_d^2$, say $\overline{\ve \mu}$, 
$\overline{\ve \sigma^2_1}, \ldots, \overline{\ve \sigma^2_d}$, respectively.  From (\ref{e-step-3}),
we obtain
$$
\overline{\ve \mu} = \frac{1}{\sum_{j=1}^n a_j} \sum_{i=1}^n {\ve x}_i
$$
and
$$ 
\overline{\ve \sigma^2_k} = \frac{1}{n}
\left [\sum_{i=1}^n b_i {\ve x}_i {\ve x}_i^T - \sum_{i=1}^n \left (\overline{\ve \mu} {\ve x}_i^T + 
{\ve x}_i \overline{\ve \mu}^T \right ) + \left ( \sum_{i=1}^n a_i \right ) \overline{\ve \mu} \ 
\overline{\ve \mu}^T 
\right ]_k.
$$
Here for a square matrix ${\ve A}$, ${\ve A}_k$ denotes the $k$-th diagonal element of the matrix ${\ve A}$.  Under the 
null hypothesis
$$
T_3 = 2(l(\widehat{p}, \widehat{\ve \mu}, \widehat{\ve \Sigma}) - l(\widehat{p}_H, \widehat{\ve \mu}_H, 
\hbox{diag}\{\widehat{\sigma}_{1H}^2, \ldots, \widehat{\sigma}_{dH}^2\})) \longrightarrow
\chi_{d(d+1)/2}^2.    
$$

\section{\sc Simulations and Data Analysis}   \label{sda}

In this section we perform some Monte Carlo simulations to show how the proposed EM algorithm performs and we 
perform the analyses of two data sets analysis to show how the proposed model and the methods can be used in practice.

\subsection{\sc Simulation Results}

For simulation purposes we have used the following sample size and the parameter values;
$$
n = 100, \ \ d = 4, \ \ p = 0.50, \ \ p = 0.75, \ \ {\ve \mu} = \left [ \matrix{0 \cr 0 \cr 1 \cr 1 \cr} \right ], \ \ {\ve \Sigma} = \left [ \matrix{2 & 2 & 1 & 0  \cr 2 & 3 & 2 & 0 \cr
1 & 2 & 3 & 2 \cr 0 & 1 & 2 & 2 \cr} \right ].
$$ 
Now to show the effectiveness of the EM algorithm we have considered both the cases namely when (a) $p$ is known and (b) $p$ is unknown.  We 
have generated samples from the above configuration and computed the MLEs of ${\ve \mu}$ and ${\ve \Sigma}$ using EM algorithm.  
In all the cases we have used the sample mean and the sample variance covariance matrix as the initial guesses of the EM algorithm.
We replicate the 
process 1000 times and report the average estimates and the associated mean squared errors (MSEs).  For known $p$, the results are 
reported in Tables \ref{kp-1} and \ref{kp-2} and for unknown $p$, the results are reported in Tables \ref{up-1} and \ref{up-2}.  In each 
box of a table, the first figure, second figure and the third figure represent the true value, the average estimate and the corresponding 
MSE, respectively.  

It is clear that the performances of the proposed EM algorithm are quite satisfactory.  It is observed that the sample mean and the sample
variance covariance matrix can be used as good initial guesses of the EM algorithm.  In all the cases considered it is observed that the
EM algorithm converges within 30 iterations, hence it can be used in practice quite conveniently.  Further, it is observed that the 
profile likelihood method is also quite effective in estimating $p$, when it is unknown.

\begin{table}
\caption{Average estimates and MSEs of $\widehat{\ve \mu}$ and $\widehat{\ve \Sigma}$, when $p$ = 0.5 and it is known}     \label{kp-1}
\bc
\begin{tabular}{|c|c|c|c|c|}   \hline
            & 0.0000 & 0.0000 & 1.0000 & 1.0000 \\
${\ve \mu}$ & 0.0053 & 0.0047 & 1.0097 & 1.0055  \\
      & (0.0984) & (0.1213) & (0.1430) & (0.1237) \\  \hline    \hline
   & 2.0000 & 2.0000 & 1.0000 & 0.0000 \\
   & 2.0024 & 1.9984 & 0.9942 & -0.0060 \\
   & (0.3299) & (0.3602) & (0.2879) & (0.2262)  \\  \cline{2-5}
   & 2.0000 & 3.0000 & 2.0000 & 1.0000  \\
   & 1.9984 & 2.9922 & 1.9907 & 0.9907 \\
$\sigma_{ij}$    & (0.3602) & (0.4932) & (0.4130) & (0.3092) \\  \cline{2-5}
               & 1.0000 & 2.0000 & 3.0000 & 2.0000  \\
  & 0.9942 & 1.9907 & 2.9757 & 1.9823 \\
   & (0.2879) & (0.4130) & (0.5350) & (0.4054)  \\ \cline{2-5}
   & 0.0000 & 1.0000 & 2.0000 & 2.0000 \\
   & -0.0060 & 0.9907 & 1.9823 & 1.9859 \\
   & (0.2262) & (0.3092) & (0.4054) & (0.3675)  \\  \hline

\end{tabular}
\ec
\end{table}

\begin{table}
\caption{Average estimates and MSEs of $\widehat{\ve \mu}$, $\widehat{\ve \Sigma}$ and $\widehat{p}$, when 
$p$ = 0.5 and it is unknown}     \label{up-1}
\bc
\begin{tabular}{|c|c|c|c|c|}   \hline
            & 0.0000 & 0.0000 & 1.0000 & 1.0000 \\
${\ve \mu}$ &-0.0067 &-0.0079 & 1.0094 & 1.0122  \\
      & (0.1029) & (0.1255) & (0.1553) & (0.1390) \\  \hline    \hline
   & 2.0000 & 2.0000 & 1.0000 & 0.0000 \\
   & 2.0105 & 2.0103 & 1.0055 & 0.0011 \\
   & (0.3579) & (0.3964) & (0.3165) & (0.2377)  \\  \cline{2-5}
   & 2.0000 & 3.0000 & 2.0000 & 1.0000  \\
   & 2.0103 & 3.0194 & 2.0095 & 1.0091 \\
 $\sigma_{ij}$   & (0.3964) & (0.5420) & (0.4443) & (0.3233) \\  \cline{2-5}
               & 1.0000 & 2.0000 & 3.0000 & 2.0000  \\
  & 1.0055 & 2.0095 & 2.9987 & 2.0006 \\
   & (0.3165) & (0.4443) & (0.5577) & (0.4161)  \\ \cline{2-5}
   & 0.0000 & 1.0000 & 2.0000 & 2.0000 \\
   & 0.0011 & 1.0091 & 2.0006 & 2.0058 \\
   & (0.2377) & (0.3233) & (0.4161) & (0.3827)  \\  \hline
\multirow{2}{*}{$p$}     & \multicolumn{4}{c|}{0.5000}   \\ 
 & \multicolumn{4}{c|}{0.5068}  \\  
     & \multicolumn{4}{c|}{(0.0433)}   \\   \hline
\end{tabular}
\ec
\end{table}

\begin{table}
\caption{Average estimates and MSEs of $\widehat{\ve \mu}$ and $\widehat{\ve \Sigma}$, when $p$ = 0.75 and it is known}     \label{kp-2}
\bc
\begin{tabular}{|c|c|c|c|c|}   \hline
            & 0.0000 & 0.0000 & 1.0000 & 1.0000 \\
${\ve \mu}$ & 0.0057 & 0.0046 & 1.0118 & 1.0068  \\
      & (0.1205) & (0.1481) & (0.1605) & (0.1350) \\  \hline    \hline
   & 2.0000 & 2.0000 & 1.0000 & 0.0000 \\
   & 2.0015 & 1.9969 & 0.9923 & -0.0065 \\
   & (0.3126) & (0.3416) & (0.2720) & (0.2099)  \\  \cline{2-5}
   & 2.0000 & 3.0000 & 2.0000 & 1.0000  \\
   & 1.9969 & 2.9906 & 1.9887 & 0.9898 \\
 $\sigma_{ij}$   & (0.3416) & (0.4645) & (0.3875) & (0.2907) \\  \cline{2-5}
               & 1.0000 & 2.0000 & 3.0000 & 2.0000  \\
  & 0.9923 & 1.9897 & 2.9778 & 1.9864 \\
   & (0.2720) & (0.3875) & (0.4848) & (0.3864)  \\ \cline{2-5}
   & 0.0000 & 1.0000 & 2.0000 & 2.0000 \\
   & -0.0065 & 0.9898 & 1.9864 & 1.9903 \\
   & (0.2099) & (0.2907) & (0.3684) & (0.3320)  \\  \hline

\end{tabular}
\ec
\end{table}

\begin{table}
\caption{Average estimates and MSEs of $\widehat{\ve \mu}$, $\widehat{\ve \Sigma}$ and $\widehat{p}$, when 
$p$ = 0.75 and it is unknown}     \label{up-2}
\bc
\begin{tabular}{|c|c|c|c|c|}   \hline
            & 0.0000 & 0.0000 & 1.0000 & 1.0000 \\
${\ve \mu}$ &0.0047 &0.0035 & 1.0141 & 1.0093  \\
      & (0.1392) & (0.1704) & (0.1938) & (0.1615) \\  \hline    \hline
   & 2.0000 & 2.0000 & 1.0000 & 0.0000 \\
   & 2.0183 & 2.0258 & 1.0263 & 0.0121 \\
   & (0.3793) & (0.4260) & (0.3453) & (0.2620)  \\  \cline{2-5}
   & 2.0000 & 3.0000 & 2.0000 & 1.0000  \\
   & 2.0258 & 3.0351 & 2.0272 & 1.0088 \\
   & (0.4260) & (0.5807) & (0.4842) & (0.3448) \\  \cline{2-5}
               & 1.0000 & 2.0000 & 3.0000 & 2.0000  \\
$\sigma_{ij}$ & 1.0263 & 2.0272 & 3.0120 & 1.9961 \\
   & (0.3453) & (0.4842) & (0.5897) & (0.4256)  \\ \cline{2-5}
   & 0.0000 & 1.0000 & 2.0000 & 2.0000 \\
   & 0.0121 & 1.0088 & 1.9961 & 1.9892 \\
   & (0.2620) & (0.3448) & (0.4256) & (0.3726)  \\  \hline
\multirow{2}{*}{$p$}     & \multicolumn{4}{c|}{0.7500}   \\ 
 & \multicolumn{4}{c|}{0.7575}  \\  
     & \multicolumn{4}{c|}{(0.0446)}   \\   \hline
\end{tabular}
\ec
\end{table}

In this section we present the analysis of two data sets namely (i) one simulated data set and (ii) one real data set mainly to
illustrate how the proposed EM algorithm and the other testing procedures can be used in practice.

\subsection{\sc Simulated Data Set}

We have generated a data set using the Algorithm 1 as suggested in Section 2, with the following specification:
$$
n = 100, \ \ d = 4, \ \ p = 0.50, \ \ {\ve \mu} = \left [ \matrix{0 \cr 0 \cr 1 \cr 1 \cr} \right ], \ \ {\ve \Sigma} = \left [ \matrix{2 & 2 & 1 & 0  \cr 2 & 3 & 2 & 0 \cr
1 & 2 & 3 & 2 \cr 0 & 1 & 2 & 2 \cr} \right ].
$$ 
It is available in http://home.iitk.ac.in/$\sim$kundu/fort.76.  We present some basic statistics of the 
data set.  The sample mean vector, and the sample variance covariance matrix are as follows:
\be
\bar{\ve x} = \left [ \matrix{
0.1489 \cr 0.1323 \cr 1.9803 \cr 1.9246 \cr} \right ] \ \ \ \hbox{and} \ \ \ {\ve S} = 
\left [ \matrix{ 
  3.4240 & 3.1869 & 1.6040 & -0.2736   \cr 
  3.1869 & 5.3792 & 4.1521 &  2.3513   \cr 
  1.6040 & 4.1521 & 7.0360 &  5.5817   \cr 
 -0.2736 & 2.3513 & 5.5817 &  6.1312   \cr} 
\right ].     \label{mvcm}
\ee
We start the EM algorithm for each $p$ with the above initial guesses.  The profile log-likelihood function is plotted 
in Figure \ref{prof-ll-1}.
\begin{figure}[h]
\bc
\epsfig{file=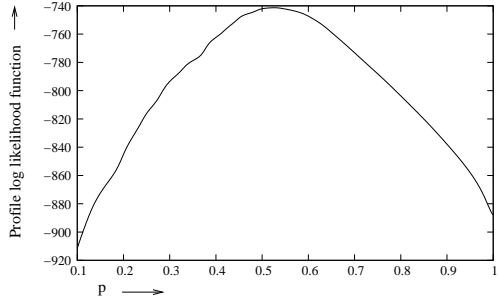, scale = 0.5}
\caption{The profile log-likelihood function.
    \label{prof-ll-1}}
\ec
\end{figure}
Finally, the MLEs of the unknown parameters are obtained as follows:
$$
\widehat{p} = 0.5260, \ \ \widehat{\ve \mu} = \left [ \matrix{0.0674 \cr 0.0636 \cr 0.9978 \cr 0.9674 \cr} \right ], \ \ \ \widehat{\ve \Sigma} = \left [ 
\matrix{
1.6511 & 1.5361 & 0.6973 & -0.2077  \cr
1.5361 & 2.5957 & 1.9258 &  1.0580 \cr 
0.6973 & 1.9258 & 2.0314 &  1.3745  \cr 
-0.2077 & 1.0580 & 1.3745 & 1.6847  \cr}  \right ], 
$$
and the associated log-likelihood value is -741.347.  It may be mentioned for each $p$, the EM algorithm is continued for 
20 iterations, and the log-likelihood value (\ref{ll-mgsn}) is calculated based on the first 50 terms of the infinite series.  
The program is written in FORTRAN-77, and it is available in http://home.iitk.ac.in/$\sim$kundu/mv-geo-sn-em-punknown-data.for.

For illustrative purposes, we would like to perform the test:
$$
H_0: p = 1, \ \ \ \hbox{vs.} \ \ \ H_1: p < 1.
$$
Under $H_0$, the MLEs of ${\ve \mu}$ and ${\ve \Sigma}$ become $\bar{\ve x}$ and ${\ve S}$, respectively, as given in 
(\ref{mvcm}), and the associated log-likelihood value is -887.852.  Therefore, the value of the test statistic $T_1$ = 
293.01, and the associated $p$ value is less than 0.00001.  Hence, we reject $H_0$.  Next we consider the following testing problem 
$$
H_0: {\ve \mu} = {\ve 0}, \ \ \ \hbox{vs.} \ \ \ H_1: {\ve \mu} \ne {\ve 0}.
$$
In this case under $H_0$, the MLEs of $p$ and ${\ve \Sigma}$ are as follows
$$
\widehat{p}_H = 0.459, \ \ \ \ \ \ 
\widehat{\ve \Sigma}_H = \left [ \matrix{ 
  1.9295 & 1.7953 & 1.0631 & 0.0073  \cr 
  1.7953 & 3.0215 & 2.4714 & 1.4589  \cr 
  1.0631 & 2.4713 & 6.1348 & 5.2589  \cr 
  0.0073 & 1.4589 & 5.2589 & 5.5066  \cr}  \right ], 
$$
and the associated log-likelihood value is -917.674.  In this case the value of the test statistics $T_2$ = 352.654.  Since
the associated $p$ value is less than 0.00001, we reject the null hypothesis.  Finally we consider the testing problem:
$$
H_0: {\ve \Sigma} = \left [ \matrix{\sigma_1^2 & 0 & 0 & 0 \cr 0 & \sigma_2^2 & 0 & 0 \cr 0 & 0 & \sigma_3^2 & 0 \cr
0 & 0 & 0 & \sigma_4^2 \cr} \right ] \ \ \ \hbox{vs.} \ \ \ H_1: {\ve \Sigma} \hbox{   is arbitrary}.
$$
In this case under $H_0$, the MLEs of the unknown parameters are as follows:
$$
\widehat{p}_H = 0.585, \ \ \widehat{\ve \mu}_H = (0.0479, 0.2502, 1.0202, 0.9967)^T
$$
and
$$
\widehat{\sigma}^2_{1H} = 1.5695, \ \ \ \widehat{\sigma}^2_{2H} = 2.2356, \ \ \ \ \widehat{\sigma}^2_{3H} = 1.1014, 
\ \ \ \ \widehat{\sigma}^2_{4H} =
0.7652.
$$
The associated log-likelihood value is -1036.80.  The value of $T_3$ = 590.91.  In this case also we reject $H_0$, as
the associated $p$ values is less than 0.00001.

\subsection{\sc Stiffness Data Set}

In this section we present the analysis of a real data set to show how the proposed model and the methodologies 
work in practice.  The data set represents the four different measurements of stiffness, $x_1, x_2, x_3, x_4$ of 
`Shock' and `Vibration' of each of 30 boards.  The first measurement (Shock) involves sending a shock wave down 
the board and the second measurement (Vibration) is determined while vibrating the board.  The last two measurements are 
obtained from static tests.  The data set is available in Johnson and Wichern \cite{JW:2013}.  For easy reference it is presented in Table \ref{data-2}.  
Since all the entries of the data set are non-negative, if we want to the 
fit the multivariate skew normal distribution to this data set,  the MLEs of the unknown parameters may not exist.  In fact we 
have tried to fit univariate skew-normal distribution to $x_1$ and it is observed that the likelihood function is an increasing 
function of the `tilt' parameter for fixed location and scale parameters.  Therefore, the MLEs do not exist in this case.  
It is expected the same phenomenon even for skew-$t$ 
distribution for large values of the degrees of freedom.

\begin{table}
\caption{Four different stiffness measurements of 30 boards}     \label{data-2}
\bc
\begin{tabular}{|c|c|c|c|c||c|c|c|c|c|}   \hline
No. & $x_1$ & $x_2$ & $x_3$ & $x_4$ & No. & $x_1$ & $x_2$ & $x_3$ & $x_4$    \\   
    &       &       &     &       &       &       &     &   &        \\    \hline
1  &  1889 &  1651 &  1561 &   1778  & 2  &  2403 &  2048 &  2087 &   2197  \\
3  &  2119 &  1700 &  1815 &   2222  & 4  &  1645 &  1627 &  1110 &   1533  \\
5  &  1976 &  1916 &  1614 &   1883  & 6  &  1712 &  1712 &  1439 &   1546  \\
7  &  1943 &  1685 &  1271 &   1671  & 8  &  2104 &  1820 &  1717 &   1874  \\
9  &  2983 &  2794 &  2412 &   2581  & 10 &  1745 &  1600 &  1348 &   1508  \\
11 &  1710 &  1591 &  1518 &   1667  & 12 &  2046 &  1907 &  1627 &   1898  \\
13 &  1840 &  1841 &  1595 &   1741  & 14 &  1867 &  1685 &  1493 &   1678  \\
15 &  1859 &  1649 &  1389 &   1714  & 16 &  1954 &  2149 &  1180 &   1281  \\
17 &  1325 &  1170 &  1002 &   1176  & 18 &  1419 &  1371 &  1251 &   1308  \\
19 &  1828 &  1634 &  1602 &   1755  & 20 &  1725 &  1594 &  1313 &   1646  \\
21 &  2276 &  2189 &  1547 &   2111  & 22 &  1899 &  1614 &  1422 &   1477  \\
23 &  1633 &  1513 &  1290 &   1516  & 24 &  2061 &  1867 &  1646 &   2037  \\
25 &  1856 &  1493 &  1356 &   1533  & 26 &  1727 &  1412 &  1238 &   1469  \\
27 &  2168 &  1896 &  1701 &   1834  & 28 &  1655 &  1675 &  1414 &   1597  \\
29 &  2326 &  2301 &  2065 &   2234  & 30 &  1490 &  1382 &  1214 &   1284  \\
\hline
\end{tabular}
\ec
\end{table}

Before progressing further we have divided all the measurements by 100, and it is not going to make any difference 
in the inferential procedure.  The sample mean vector and the sample variance covariance matrix of the transformed data are
\be
\bar{\ve x} = \left [ \matrix{19.0610 \cr 17.4953  \cr 15.0790  \cr 17.2497  \cr} \right ] \ \ \ \hbox{and} \ \ \ 
{\ve S} = \left [ \matrix{
  10.2096 & 9.1460 & 8.4590 & 9.1090 \cr
   9.1460 & 9.8126 & 7.3791 & 7.8362 \cr 
   8.4590 & 7.3791 & 8.9212 & 8.7615 \cr 
   9.1090 & 7.8362 & 8.7615 &10.0754. \cr}  \right ]      \label{mv-2}
\ee
Based on the EM algorithm and using the profile likelihood method, we obtain the MLEs of the unknown parameters as follows:
$$
\widehat{p} = 0.9640, \ \ \widehat{\ve \mu} = \left [ \matrix{18.2409 \cr 16.7594  \cr 14.4459  \cr 16.4719  \cr} \right ], 
\ \ \ \widehat{\ve \Sigma} = \left [ \matrix{
  7.6625 & 6.4895 & 6.1284 & 7.4978   \cr
  6.4895 & 7.0296 & 4.9941 & 6.0798   \cr 
  6.1284 & 4.9941 & 6.7659 & 7.1631   \cr 
  7.4978 & 6.0798 & 7.1631 & 9.1815.   \cr} \right ],
$$
and the associated log-likelihood value is -271.969.  Now to check whether the proposed MGSN distribution provides a better 
fit than the multivariate normal distribution or not, we perform the following test:
$$
H_0: p = 1, \ \ \ \hbox{vs.} \ \ \ H_1: p < 1.
$$
Under $H_0$, the MLEs of ${\ve \mu}$ and ${\ve \Sigma}$ are provided in (\ref{mv-2}), and the associated log-likelihood value
is -277.761.  Therefore, the value of the test statistic $T_1$ = 11.584, and the associated $p$ value is 0.0000025.  Hence, we
reject the null hypothesis, and it 
indicates that the proposed MGSN distribution provides a better fit than the multivariate normal distribution to the given 
stiffness data set.  AIC also prefers MGSN distribution than the multivariate normal distribution for this data set.

\section{\sc Conclusion}

In this paper we have discusses different properties of the MGSN distribution in details.  Different characterization results
and dependence properties have been established.  The $d$-dimensional MGSN distribution has $d+1+d(d+1)/2$ unknown parameters.  
We have proposed to use EM algorithm and the profile likelihood method to compute the MLEs of the unknown parameters, and it is
observed that  the proposed algorithm can be implemented very easily.   We have discussed some testing of hypothesis problems also.  
Two data sets have been analyzed to show the effectiveness of the proposed methods, and it is observed that for the real 
'stiffness' data set MGSN provides a better fit than the multivariate normal distribution.  Hence, this model can be used as an
alternative to Azzalini's multivariate skew normal distribution.

\section*{\sc Acknowledgements} The author would like to thank Prof. Dimitris Karlis from the Athens University of 
Economics and Business, Greece, for his helpful comments.  The author is really thankful to two reviewers and the associate
editor for their careful reading and providing constructive comments.

\end{document}